# CONTROL AND ULTRAFAST DYNAMICS OF A TWO-FLUID POLARITON SWITCH

M. De Giorgi[1,2], D. Ballarini[2], E. Cancellieri[3], F. M. Marchetti[3], M. H. Szymanska[4,5], C. Tejedor[3], R. Cingolani[6], E. Giacobino[7], A. Bramati[7], G. Gigli[1,2,8], and D. Sanvitto[1,2]

[1]NNL, Istituto Nanoscienze - CNR, Via Arnesano, 73100 Lecce, Italy

[2]Istituto Italiano di Tecnologia, IIT-Lecce, Via Barsanti, 73010 Lecce, Italy

[3] Física Teórica de la Materia Condensada, Universidad Autónoma de Madrid, Spain

[4]Department of Physics, University of Warwick, Coventry, England

[5]London Centre for Nanotechnology, London, UK

[6]Istituto Italiano di Tecnologia, IIT-Genova, Genova, Italy

[7]Laboratoire Kastler Brossel, Université Pierre et Marie Curie-Paris 6, École Normale Supérieure et CNRS, UPMC Case 74, 4 place Jussieu, 75005 Paris, France

[8]Innovation Engineering Department, University of Salento, Via Arnesano, 73100 Lecce, Italy

**We investigate the cross-interactions in a two-component polariton quantum fluid coherently driven by two independent pumping lasers tuned at different energies and momenta. We show that both the hysteresis cycles and the ON/OFF threshold of one polariton signal can be entirely controlled by a second polariton fluid. Furthermore, we study the ultrafast switching dynamics of a driven polariton state, demonstrating the ability to control the polariton population with an external laser pulse, in less than a few picoseconds**.

Although electronic technology has dominated a wide share of the market for communication and computation, becoming extremely sophisticated and well studied, it seems now unable to keep up with the same trend given the strong demand for fast networking performances and low energy consumption. Recent progresses on photonic integrated circuits promise to overcome the limits of conventional electronic technology [1-5], offering the opportunity to realize efficient signal processing at speeds much higher than in conventional electronic devices, with further advantages for low energy consumption and negligible heating. Optical switching is an essential component of optical communication networks and a fundamental milestone for quantum computation, allowing one optical signal to be controlled by another photonic beam. Various mechanisms have been proposed to achieve all-optical switching devices such as photonic-bandgap shift [6,7] and defect-mode shift [8, 9]. Here, the refractive-index contrast of the material is modified via a $\chi^2$ non-linearity by an external optical pump, which, however, requires high intensities (often of the order of GW/cm$^2$) to achieve large switching efficiencies [8,10-11]. Another approach is the use of materials with large nonlinear optical coefficients but with slower non-linear response time [12]. Recently, a proof of principle demonstration of excitonic switching devices was reported [13]. The advantage of using excitons is their strong $\chi^3$ non-linearities, with the drawback of being limited by their very slow intrinsic lifetime. On the other hand, microcavity polaritons, the strongly coupled quasiparticles between a quantum well exciton and a cavity photon, are especially interesting and promising systems to study, since they combine the properties of photons with the high non-linearity of excitons. As bosonic quasiparticles, polaritons have unique coherent properties that have led to the achievement of Bose–Einstein condensation and superfluidity [14, 15, 16]. In addition, polariton-polariton interaction [17,18] is at the basis of phenomena such as optical bistabilities [19, 20], OPO [21], spin dependent activation of the bistable behaviour [22, 23], polariton switch operations via an additional non-resonant laser [24], and, more recently, transistor devices [25]. The polariton system is thus the ideal candidate for high-speed operations in logic devices at low energy consumption.

Optical bistability in microcavities is the enhanced resonant absorption of a detuned optical laser pump because of polariton interactions [26]:at low laser powers, by tuning the pump above the polariton dispersion, the driven polariton state is almost empty (OFF state). Increasing the power, the polariton blue-shift due to interactions causes enhanced absorption, a superlinear growth of the polariton population, and, eventually, a jump into a high polariton density (ON state). Theemission intensity versus the pump power is characterized by a hysteresis loop, which strongly depends on the excitation conditions [27]. In the case of two pumping lasers, however, the mutual interactions between the two driven polariton states, imply a richer phenomenology.

In this paper, we investigate a two-component polariton system coherently driven by two different lasers with independently tunable frequencies, wavevectors and intensities. Stimulated by the theoretical analysis of Refs.[28, 29], we demonstrate the possibility of controlling the hysteresis cycles of a polariton state via another state. Further, using a combination of continuous wave (CW) and pulsed laser beams, we show the possibility of ultrafast control of a polariton state, and in particular the ability to switch a state not only 'ON' but also 'OFF' within a timescale of few picoseconds.

We use two spatially overlapping continuous wave (CW) Ti:Sapphire lasers (P1 and P2) with different k-vectors and frequencies which resonantly excitetwo polariton states in the lower polariton branch (LPB) in a GaAs/AlAs microcavity (front/back reflectors with 21/24 pairs) containing three In$_{0.04}$Ga$_{0.96}$As/GaAs quantum wells (the sample is kept at 10 K). We choose the k-vectors and frequencies of both lasers so that to ensure optical bistability for each of the two states independently (upper panel in Fig.1). Momentum space images of the emission intensity from the two states S1 and S2 at the pump energy $E_1$ and $E_2$ and momentum of $k_1$ and $k_2$, respectively, were detected in transmission geometry by using a high-resolution imaging CCD camera coupled to a spectrometer.

At the same time, we theoretically simulate the experimental system by means of a generalized Gross-Pitaevskii equation for the cavity $\Psi_C$ and exciton $\Psi_X$ fields ($\hbar=1$):

$$i\partial_t \begin{pmatrix} \psi_X \\ \psi_C \end{pmatrix} = \begin{pmatrix} 0 \\ F \end{pmatrix} + \begin{pmatrix} \omega_X - i\kappa_X + g_X|\psi_X|]^2 & \frac{\Omega_R}{2} \\ \frac{\Omega_R}{2} & \omega_C - \frac{\nabla^2}{2m_C} - i\kappa_C \end{pmatrix} \begin{pmatrix} \psi_X \\ \psi_C \end{pmatrix}. \quad (1)$$

The fields are coupled by the Rabi splitting $\Omega_R$ and resonantly driven by the two external lasers $F = f_1(r) e^{i(k_1 r - E_1 t)} + f_2(r) e^{i(k_2 r - E_2 t)}$ having frequencies and momenta close to the experimental ones ($E_1 = -4.25$ meV, $E_2 = -3.7$ meV, measured with respect to $\omega_X$, $k_1 = 0.2$ µm$^{-1}$, $k_2 = 0.3$ µm$^{-1}$). Here, $\kappa_C = 0.1$ meV and $\kappa_X = 0.001$ meV are the photon and exciton decay rates, $m_C = 2 \times 10^{-5} m_0$ the photon mass ($m_0$ is the electron mass), the exciton interaction strength $g_X$ is set to one by rescaling both fields $\Psi_{C,X}$ and pump intensities $f_{1,2}$. The method used to solve Eq.(1) is the same one employed in Ref.[28], thus we give here only a short account of it. We establish, within the linear response approximation and for homogeneous pumps ($f_{1,2}(r)=f_{1,2}$), the stability of the system as a function of the two pump intensities. In other words, we first solve the Gross-Pitaevskii equation at the mean-field level, and then, we establish the dynamical stability of each solution to small fluctuations. In this regime we approximate the mean-field solutions of the system with $\Psi_{C,X}(r,t) = \Sigma_{j=1,2} \Psi^{ss}_{jC,X} e^{i(k_j r - E_j t)}$, where $|\Psi^{ss}_{jC,X}|^2$ are the photon and exciton populations at frequency $E_j$ and momentum $k_j$.

For the system parameters specified above, we plot the stability phase diagram for the two pumped states as a function of the two rescaled pump powers $f_1'$ and $f_2'$ in Fig. 1. The colour scheme means that the state S1 on the left panel and S2 on the right panel are weakly populated (OFF) in the yellow region; two solutions either weakly (OFF) or strongly (ON) populated coexist in the two bistable regions in green; finally, in the red region only the strongly populated (ON) solution is stable. It is interesting to note that the threshold values between the ON and the OFF states of one fluid can strongly depend on the intensity of the pump of the other state.

To better understand the behaviour of the system, we compare, in Fig. 2, the experimental (1e, 2e, 3e and 4e) and theoretical (1t, 2t, 3t and 4t) emission intensities from the two states by either fixing the pump power P2 and changing P1 (left column of Fig. 2 and horizontal black lines in the phase diagram of Fig. 1), or by fixing P1 and changing P2 (right column of Fig. 2 and vertical black lines in the phase diagram of Fig. 1). We choose the values of the fixed pump powers so that to represent the most different configurations for the initial and final state. In both experimental and theoretical plots, the emission intensities are normalized to the maximum values of either the S1 or S2 state when they are excited independently.

In two cases (1 and 2) we start with the dressed LPB red-detuned with respect to both lasers, while in the other two cases (3 and 4) we start with the LPB blue-detuned with respect to laser 1 but red-detuned with respect to laser 2. As it can be observed from Fig.2, by controlling the power of one pump, and therefore the population of the corresponding state (the hysteresis of which is shown with solid lines) we are able to control the population of the other polariton state (solid and open symbols for increasing and decreasing pump power, respectively).

In the first case (panels 1e and 1t), P2 is fixed below the hysteresis cycle threshold of S2 when P1=0. Here, the LPB is red-detuned with respect to both pumping lasers. By increasing P1, the LPB becomes resonant with $E_2$ and is therefore pushed up far above $E_1$. In these conditions, the system is filled with polaritons in S2 but with just few polaritons in S1, even when P1 reaches values above the threshold it would have when P2 = 0. When P1 is again reduced to zero, the intensity of P2 is not sufficient to keep the LPB blue-detuned and therefore the system goes back to the initial empty state. This demonstrates that a full hysteresis cycle of the state S2 can be completely controlled by the pump P1.

In the second case (panels 2e and 2t), the initial conditions are similar to the previous case. Here, however, P2 is varied and P1 is kept constant to a value large enough to sustain the LPB in the ON state of S1. When P2 is turned on, the LPB enters again in resonance with $E_2$ and is therefore far above $E_1$ and the cavity is filled with polaritons in S2. In this case, however, when P2 is decreased, even if the LPB red-detunes to values smaller than $E_2$, it is sustained in resonance with $E_1$ by P1 and therefore the cavity remains now filled with polaritons in S1. The net effect of this cycle is that P2 can be used to turn ON the state S1.

In the third case (panels 3e and 3t), the LPB is blue-detuned by P2, so that is higher than $E_1$ and lower than $E_2$. The value of P2 is chosen in order to be strong enough to sustain the ON state of S2. When P1 is increased, it pushes the LPB in resonance with $E_2$ and S2 turns ON while S1 remains poorly populated. When P1 is decreased to zero, the LPB is sustained in resonance with $E_2$ by P2. The net effect of this cycles is the opposite of the previous one, a cycle in the intensities of P1 turns the state S2 ON.

Finally, in the fourth case (panels 4e and 4t), the LPB is blue-detuned so that to be higher than $E_1$ and lower than $E_2$,. When P2 is turned on, the LPB enters in resonance with $E_2$, it fills with polaritons in state S2 and stay far above $E_1$, so that the population in S1 decreases almost to zero. When P2 is turned off again, the LPB red-detunes and goes back in resonance with $E_1$, and the system goes back to its initial conditions. As a consequence, the S1 state is reversibly switched ON and OFF by turning OFF and ON a different S2 state.

This allows to control not only the ON, but also the OFF state of a polariton quantum fluid via another polariton state.

From this analysis it emerges that, for two interacting polariton fluids, one polariton state can be used to control the population of the other state. Depending on both the relative intensities and the relative detuning of the two pumps, the system can be brought *in* and *out* of resonance with the pump frequencies, and so the two states can be turned ON and OFF. As a final remark, we would like to address some differences between the theoretical curves and the experimental data in Fig. 2. While the theoretical curves present sharp transitions between the ON and OFF states, as well as extremely low population in the OFF states, the experimental curves display smoother crossover and slightly populated OFF states. These differences can be in part ascribed to temperature fluctuations and time averaging over different realization. Furthermore, one has to take into account that, while the theoretical analysis is carried on with delta-like laser-lines, in the experiments at least one of the two CW lasers was not a monomode laser, allowing for an effective broader excitation line, which justifies the observation of intermediate conditions. For the same reason, the bistable region of S2 (see zoom in Fig. 1) depends on the pump power $f_1$' in a much weaker way in the theoretical analysis than in the experimental case.

In order to measure the ON/OFF switching time of the state S1, we pump the system in the state S2 with a Ti:Sapphire pulsed laser with a pulse width of 120 fs and a repetition rate of 82 MHz, while the state S1 is populated by the monomode CW laser with linewidth < 5 MHz. This has been performed on a different point of the sample with a more positive detuning, where the ON/OFF switch effect is more pronounced. Energies and momenta of the two lasers are chosen as shown in the inset of Fig. 3(a): due to the wide spectral range of the pulsed laser, P2 is always on resonance with the LPB, whereas E and k of the CW laser P1 has been chosen so that to avoid optical bistability for S1 when P2=0. Time resolved photoluminescence of the S1 state is performed in transmission geometry by using a Hamamatsu streak camera coupled to a 0.55 m spectrometer (time resolution 5 ps).

The effect of the pulsed laser on the state S1 is shown in Fig. 3 for different excitation conditions. Below threshold, when $E_1$ is still slightly detuned above the LPB energy, S1 is OFF [black line in the inset of Fig. 3(b)]. The pulsed laser fast induces a blue-shift [red dashed line in the inset of Fig. 3(b)], leading to a strong increase of the polariton population in the state S1. This results in a switch ON of the S1 state [Fig. 3(b)]. Once the pulse is gone away, the S1 state comes back to its original steady state, with a recovery time of hundreds of picoseconds. Conversely, above the P1 power threshold [Fig. 3(c), 3(d)], when S1 is already in the ON state [black line in the inset of Fig. 3(d)], the further blue-shift induced by the pulse laser (red dashed line) brings the S1 state out of resonance, resulting in a rapid reduction of the polariton population to less than ½ of its original density (OFF state). Similarly to the previous case, the S1 polariton population returns to the initial condition in many hundreds of ps. This effect is further enhanced for higher power of the pulsed pump, resulting in a bigger variation of the polariton population between the ON and OFF states [Fig. 3(d)], and in a longer recovery time. Such long recovery time could be ascribed to a dynamical parametric instability, which is temporarily triggered by the ultrafast (and thus broad in energy) laser, similarly to what it has already been observed in recent TOPO experiments [30]. In all cases, we estimate a switching ON and OFF times for the S1 signal state of 5 ps, which is limited by the time resolution of the system [not visible in the long time range used for Fig. 3].

To conclude, we have studied the stability of a two-component polariton quantum fluid coherently driven by two independent pump lasers. We clearly demonstrate that we are able to control the hysteresis loop of a polariton quantum fluid by changing the population of a second polariton state and show how the polariton non-linear optical properties can be used to switch not only ON but also OFF a polariton fluid. Finally, we have observed a very short, of the order of a few picoseconds, switching time between the ON and OFF state, establishing microcavity polaritons as promising systems for ultrafast optical operations.

**Acknowledgements**

The authors acknowledge P. Cazzato for the technical assistance with the experiments. This work has been partially funded by the FIRB Italnanonet, FIRB Italy-Japan HUB on nanotechnologies, the POLATOM ESF Research Networking Program, the Spanish MINECO (MAT2011-22997), CAM (S-2009/ESP-1503), and the program Ramon Y Cajal (F.M.M.).

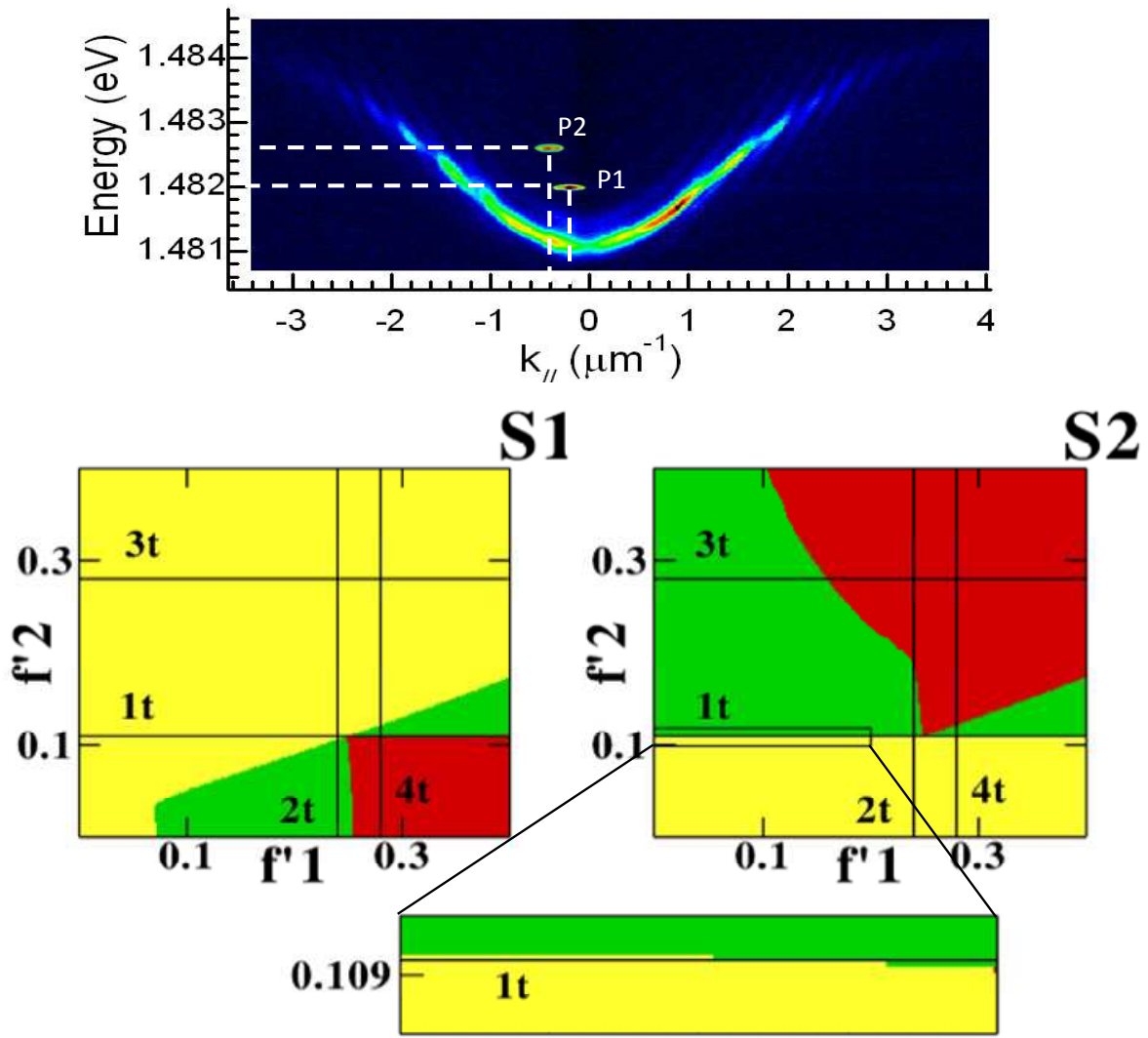

**Figure 1.** Upper panel: experimental polariton dispersion relation with superimposed the two laser pumps at the corresponding energy ($E_1$ = 1.482 eV and $E_2$ = 1.4826 eV) and momenta ($k_1$ = 0.2 µm$^{-1}$ and $k_2$ = 0.4 µm$^{-1}$) used to excite two different polariton states. Lower panels: theoretically evaluated phase diagram showing the OFF and ON states of S1 (left panel) and S2 (right panel) as a function of the rescaled pump intensities $f_{1,2}' = \sqrt{g_X} f_{1,2}$ [meV$^{3/2}$]. In the yellow region the state Si can only be weakly populated (OFF), while in the red one can only be strongly populated (ON). In the green region both ON and OFF solutions coexist (bistable region). The black thick vertical and horizontal lines correspond to the cases studied in Fig 2, where one pump is kept constant while the other scans different intensity values.

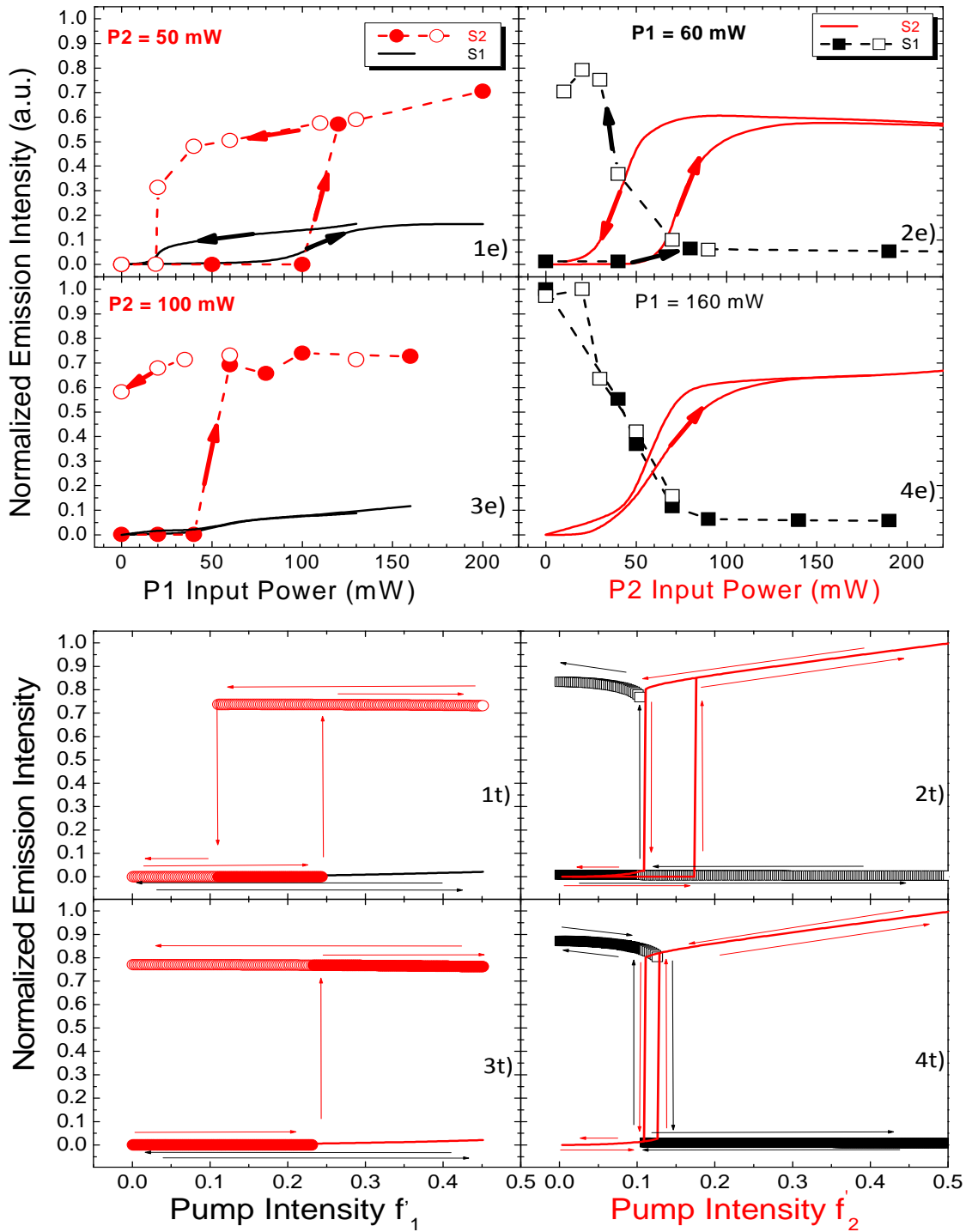

**Figure 2.** Experimental (1e, 2e, 3e, 4e) and theoretical (1t, 2t, 3t and 4t) normalized emission intensities from the two states S1(black) and S2(red) obtained by either changing the pump power P1 for fixed P2 (left column), or changing P2 for a fixed P1 (right column). Arrows and symbols correspond to the power change variation: increasing power, solid symbol, decreasing power open symbols. Note that for the theoretical emission intensity we consider $|\Psi^{SS}_{1C}|^2$ and $|\Psi^{SS}_{2C}|^2$ respectively, which are proportional to the experimental emission intensities when the Hopfield factors are taken into account.

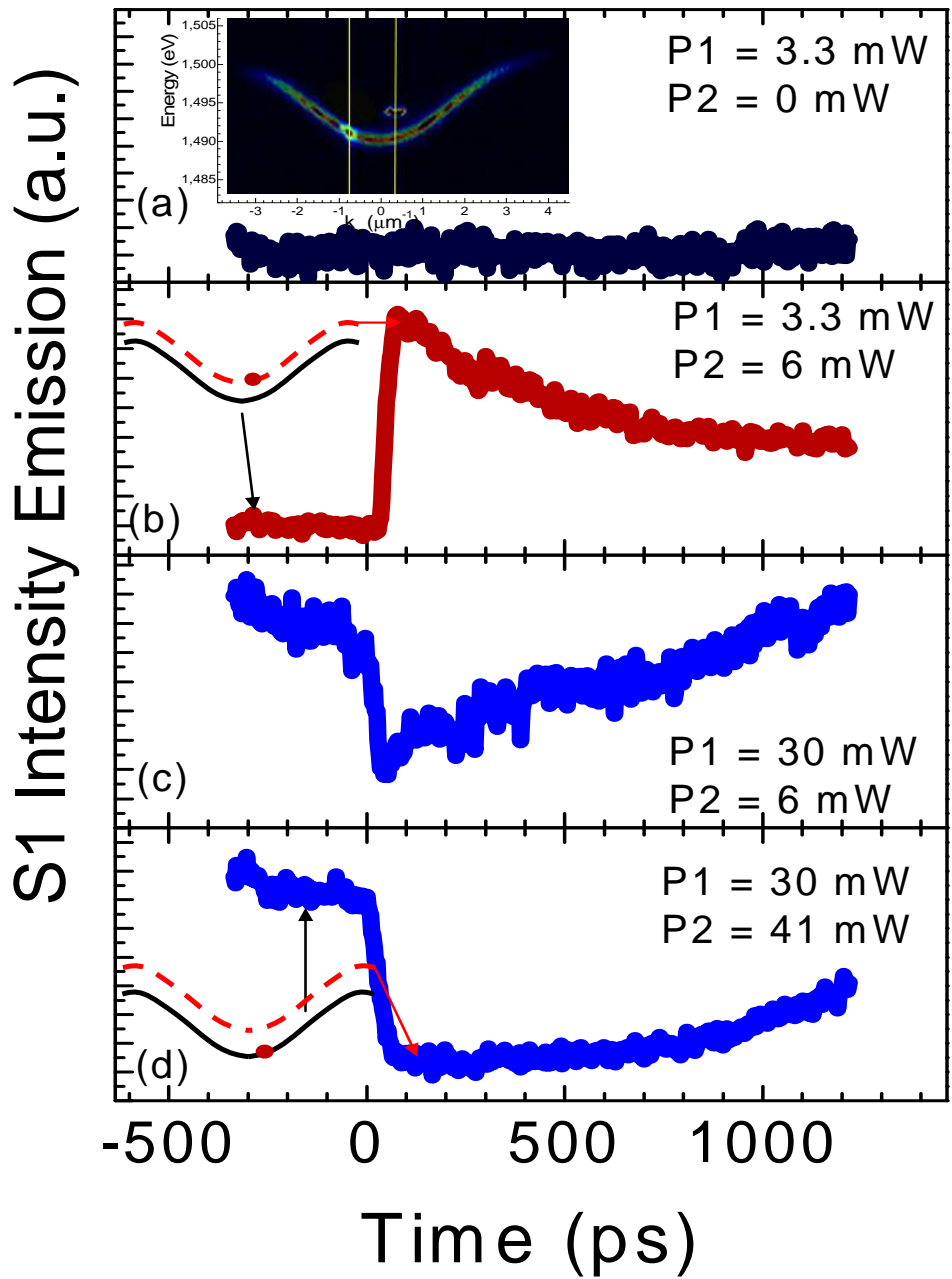

**Figure 3:** Time resolved photoluminescence intensity of the state S1 excited by a CW (P1) and a fs pulsed (P2) laser under the conditions shown in the inset of Fig. 2a: k1 = 0.35 µm$^{-1}$ and E1 = 1.494 eV, k2 = 0.73 µm$^{-1}$ and E2 = 1.491 eV. The spectra shown in the figure are obtained under four different excitation conditions: (a) the state is pumped only by the CW laser P1 at a very low pump power of 3.3 mW. Under this condition S1 is on an OFF state; (b) S1, same as in (a) but a fs pulsed laser exciting the system with a low power P2=6 mW switches ON the S1 state; (c-d) S1, in an ON state (continuously pumped by P1 at 30 mW), is switched OFF by the pulsed laser. The inset in (b) and (d) shows the influence of a fs pulsed laser (P2) pumping the system resonantly. The continuous pump P1 is schematized by the red dots.